\documentclass[fleqn,usenatbib]{mnras}

\usepackage{newtxtext,newtxmath}

\usepackage[T1]{fontenc}
\usepackage{ae,aecompl}

\usepackage{graphicx}	
\usepackage{amsmath}	
\usepackage{amssymb}	

\usepackage{threeparttable}
\usepackage{float}
\usepackage{epstopdf}

\title[Dust Attenuation of Star-Forming Galaxies at $z=2.5-4.0$]{A$^3$COSMOS: The Dust Attenuation of Star-Forming Galaxies at $z=2.5-4.0$ from the COSMOS-ALMA Archive }

\author[Y. Fudamoto et al.]{
Yoshinobu Fudamoto$^{1}$\thanks{E-mail: yoshinobu.fudamoto@unige.ch},
P. A. Oesch$^{1,2}$,
B. Magnelli$^{3}$,
E. Schinnerer$^{4}$,
D. Liu$^{4}$,
\newauthor
P. Lang$^{4}$,
E. F. Jim{\'e}nez-Andrade$^{3}$,
B. Groves$^{5}$,
S. Leslie$^{4}$,
M. T. Sargent$^{6}$,
\newauthor
\\
$^{1}$Department of Astronomy, University of Geneva, 51 Ch. des Maillettes, 1290 Versoix, Switzerland\\
$^{2}$International Associate, Cosmic Dawn Center (DAWN), Copenhagen, Denmark\\
$^{3}$Argelander-Institut f\"ur Astronomie, Universit\"at Bonn, Auf dem H\"ugel 71, D-53121 Bonn, Germany\\
$^{4}$Max Planck Institute for Astronomy, K\"{o}nigstuhl 17, 69117 Heidelberg, Germany\\
$^{5}$Research School of Astronomy \& Astrophysics, Australian National University, Mt. Stromlo Observatory, Cotter Rd, Weston Creek, ACT,\\ Australia\\
$^{6}$Astronomy Centre, Department of Physics and Astronomy, University of Sussex, Brighton BN1 9QH, UK\\
}

\date{Accepted XXX. Received YYY; in original form ZZZ}

\pubyear{2019}

\begin{document}
\label{firstpage}
\pagerange{\pageref{firstpage}--\pageref{lastpage}}
\maketitle

\begin{abstract}
We present an analysis of the dust attenuation 
of star forming galaxies at $z=2.5-4.0$ through the relationship between the UV spectral slope ($\beta$), stellar mass ($M_{\ast}$) and the infrared excess (IRX$=L_{\rm{IR}}/L_{\rm{UV}}$) based on far-infrared continuum observations from the Atacama Large Millimeter/sub-millimeter Array (ALMA).
Our study exploits the full ALMA archive over the COSMOS field processed by the A$^3$COSMOS team, which includes an unprecedented sample of $\sim1500$ galaxies at $z\sim3$  as primary or secondary targets in ALMA band 6 or 7 observations with a median continuum sensitivity of 126 $\rm{\mu Jy/beam}$ (1$\sigma$).
The detection rate is highly mass dependent, decreasing drastically below $\log (M_{\ast}/M_{\odot})=10.5$. 
The detected galaxies show that the IRX-$\beta$ relationship of massive ($\log M_{\ast}/M_{\odot} > 10$) main sequence galaxies at $z=2.5-4.0$ 
is consistent with that of local galaxies, while starbursts are generally offset by $\sim0.5\,{\rm dex}$ to larger IRX values.
At the low mass end, we derive upper limits on the infrared luminosities through stacking of the ALMA data. The combined IRX-$M_{\ast}$ relation at $\rm{log\,(M_{\ast}/M_{\odot})>9}$ exhibits a significantly steeper slope than reported in previous studies at similar redshifts, implying little dust obscuration at $\log M_{\ast}/M_{\odot}<10$. However, our results are consistent with early measurements at $z\sim5.5$, indicating a potential redshift evolution between $z\sim2$ and $z\sim6$. Deeper observations targeting low mass galaxies will be required to confirm this finding. \\
\end{abstract}







\section{Introduction}
\label{sec:intro}
The total census of the star formation rate density (SFRD) of galaxies across cosmological time is a major milestone of  modern extragalactic research \citep[e.g. review by][and references therein]{Madau2014}.
Deep rest-frame ultra-violet (UV) photometry obtained by the Hubble Space Telescope (HST), especially after the advent of the Wide Field Camera (WFC3) has now collected large galaxy samples at $z\sim7-8$, and has even pushed the measurement of the SFRD to $z>10$ \citep[][]{Ellis2013,Mclure2013,Schenker2013,Finkelstein2015,Oesch2015,Oesch2018,Bouwens2014,Kawamata2015,Laporte2016,Mcleod2016}.
However, HST provides a very biased view of high-redshift galaxies, in particular at $z>3$ where it only has  access to the rest-frame UV.
The UV emission is highly sensitive to absorption by interstellar dust.
The absorbed energy is re-emitted as thermal emission at far-infrared (FIR) wavelengths, and dominates the galaxies' bolometric energy output until redshift $z\sim2$, and  contributes significantly out to even higher redshift
\citep{LeBorgne2009,Magdis2012,Magnelli2013,Bethermin2015,Casey2018b,Zavala2018}.
Thus, the understanding of the dust extinction and re-emission, in various environments and at different redshifts is a crucial ingredient in studying the star formation activity of galaxies, in particular in the high-redshift Universe.

The empirical relation between the infrared excess ($=L_{\rm{IR}}/L_{\rm{UV}}$; hereafter IRX), and UV spectral slope ($\beta$; $f_{\lambda}(\lambda)\propto\lambda^{\beta}$) is the most commonly used tool to derive a correction for dust attenuation \citep[][]{Meurer1999,Gordon2000}. 
In the local Universe, studies find starburst galaxies to follow a relatively tight relationship in the IRX-$\beta$ diagrams \citep[][]{Meurer1999,Takeuchi2010,Overzier2011}.
In particular, \citet[][hereafter M99]{Meurer1999} derived the IRX-$\beta$ relation using a sample of compact starburst galaxies observed by the {\it International Ultraviolet Explorer} \citep[IUE,][]{Kinney1993}, and FIR flux measurements from the {\it Infrared Astronomical Satellite}.
The M99 relation is applied to estimate the dust attenuation over virtually the entire redshift range of galaxies, with the assumption that high-redshift young galaxies share similar characteristics with  local starburst galaxies.

However, even in the local Universe, the IRX relation presented by M99 has been extensively revised by several studies.
Re-measurements of the M99 galaxies with the {\it Galaxy Evolution Explorer} (GALEX) reveal that a large part of the UV fluxes were previously underestimated due to the small aperture of IUE, and this biases the M99 relation towards higher IRX at fixed $\beta$ \citep{Overzier2011,Takeuchi2012}.
Also, it is highly debated if the IRX-$\beta$ relation is universally applicable to  local non-starburst galaxies.
While some studies find that the general galaxy population follows a relatively tight IRX-$\beta$ relation \citep{Casey2014,Salim2018}, other studies find significant scatter due to the contribution from  older stellar populations, and  offsets that might be correlated with stellar population ages \citep{Kong2004,Buat2005,Grasha2013}.

At higher redshift, the situation is even less clear.
Using IR observations from {\it Spitzer}/MIPS, early studies have found that an IRX-$\beta$ relation consistent with M99  is already established at redshift up to $z\sim2.5$, although with significant scatter \citep{Reddy2010,Reddy2012,Whitaker2012}.
Using deep ALMA observations, \citet{Bouwens2016} showed that the M99 relation is not applicable for $z>2$ less massive galaxies with $M_{\ast} < 10^{9.75}\,\rm{M_{\odot}}$, and a steeper, SMC-like dust extinction curve is required to explain the low observed infrared fluxes of galaxies in the HUDF. 
On the other hand, several studies find that the M99 relation does still hold for $z\sim2-3$ galaxies on the main sequence  \citep[][]{AlvarezMarquez2016,Fudamoto2018,Mclure2018}.
\citet{Capak2015} and \citet{Faisst2017} show that, for $z\sim5.5$ galaxies, the IRX is much lower than the M99 relation unless one assumes a significantly higher dust temperature ($T_d >50\,\rm{K}$).

The correlations between IR luminosity and other physical properties, in addition to $\beta$, have also been investigated.
Studies show that the IRX is also strongly correlated with stellar mass ($M_{\ast}$) \citep[][]{Heinis2014,Pannella2015,AlvarezMarquez2016,Dunlop2017,Koprowski2018}.
This is not surprising, as the stellar mass of a galaxy reflects its past star forming activity, and is a priori an indicator of the dustiness of the interstellar material as the dust particles are produced through stellar activity such as pulsating moderate mass stars and supernovae \citep[e.g.][]{Dwek1980}.

One major shortcoming of many previous studies was that they had to rely on {\it stacking} of low resolution images, e.g. from {\it Herschel}, to infer the average infrared luminosity of a sample of galaxies, which is significantly affected by foreground contamination and clustered, neighbouring objects. While stacking analyses can reveal the average properties, they lose important information such as the scatter in the population. Thus, it is still unclear if there is a large scatter in the IRX relations and/or if there is a fraction of galaxies that are significantly offset from the average relations. The only way to resolve previous discrepancies are higher resolution FIR continuum measurements of individual galaxies to study the scatter in the IRX-$\beta$/$M_{\ast}$ relation at high redshifts. This has become possible after the developments of sensitive/high-resolution millimetre or sub-millimetre observing facilities such as the Atacama Large Millimeter/submillimeter Array (ALMA) and the Northern Extended Millimeter Array (NOEMA).

Using these facilities, the IRX-$\beta$/$M_{\ast}$ of high-redshift star forming galaxies is now routinely studied, yet important open questions remain.
In our previous study \citep{Fudamoto2018}, we use ALMA band 6 observations of massive ($\rm{log}\,M_{\ast}/M_{\odot} > 10.5$) star forming galaxies to show that massive galaxies at $z\sim3$ generally follow the local relation.
However, the sample size ($\sim60$ galaxies) and the dynamic range of $\beta$/$M_{\ast}$ covered were small, limiting the conclusions.

In this paper, we update our previous studies on the IRX-$\beta$ and the IRX-$M_{\ast}$ relationship of star forming galaxies at $z=2.5-4.0$
using individual FIR flux measurements for 1512 galaxies (of which $\sim10\%$ are $>3\sigma$ detections) obtained from public ALMA archival data over the $2\,\rm{deg^2}$ Cosmic Evolution Survey field \citep[COSMOS;][]{Scoville2007,Koekemoer2007}, in order to provide new insight into the dust attenuation of high-redshift galaxies.
The paper is organised as follows: in \S2 we describe our sample and galaxy selection, and \S3 presents the method of our analysis.  \S4 shows the results on the IRX-$\beta$/$M_{\ast}$ relation obtained from our sample, and we conclude our study in \S5.

Throughout this paper, we assume a cosmology with $(\Omega_m,\Omega_{\lambda},h_0)=(0.3,0.7,0.7)$, and a Chabrier IMF \citep{Chabrier2003} where applicable.
When comparing to previous measurements in the literature with a \citet{Salpeter1955} IMF, we divide masses, $M_{\ast,\rm{Salpeter}}$, by 1.58 \citep{Ilbert2010,Madau2014}, and SFR$_{\rm Sapleter}$ by 1.78 \citep{Salim2007} to convert to a Chabrier IMF.

\section{Data and Sample Selection}
\label{sec:data}
In this section, we present our sample obtained from the public ALMA archive in the COSMOS field.
All the steps of our data acquisition, imaging, source extraction, and FIR continuum photometry are fully described in our data release paper \citep[A$^3$COSMOS\footnote{\url{https://sites.google.com/view/a3cosmos}};][]{Liu2018}, and interested readers are referred to it for more detail. 
Below, we only provide a brief summary.

\subsection{A$^3$COSMOS Prior Photometry Data Based on the COSMOS2015 Catalogue }
\label{sec:A3COSMOSsample}
We made use of all the ALMA archival band 6 and band 7 observations 
in the COSMOS field, that were publicly available as of January 2018.
After obtaining the raw data from the ALMA archive, all the data are calibrated using the script released by the QA2 analyst for each PI project (i.e. $\rm{scriptForPI.py}$).
These scripts use the $\rm{Common\,Astronomy\,Software\,Applications\, package}$ \citep[CASA:][]{Mcmullin2007}. In particular, we used the appropriate version of CASA  as specified by the scripts.
The calibrated visibility data are then imaged and deconvolved
from their dirty-beam with the $\rm{CLEAN}$ algorithm of the imaging pipeline using a Briggs weighting scheme with a stopping threshold of $S/N=4$ and a robustness parameter of 2 (i.e. natural weighting), and spatial extent up to a primary beam attenuation of 0.2.
Projects for which the imaging pipeline failed, were instead imaged manually by using a consistent weight and a spatial extent.

The ALMA photometry relies on a prior source extraction. We only selected galaxies that have counterparts in the deep optical to near-infrared photometry catalogue of the COSMOS field \citep[hereafter COSMOS2015:][]{Laigle2016}.
The COSMOS2015 catalogue is based on the source extraction on a combined NIR image from the UltraVISTA survey \citep[J-,H-,Ks- band;][]{McCracken2012}, and the z$^{++}$ image from the Subaru telescope.

The photometry from the ALMA images is obtained using GALFIT   \citep{Peng2002,Peng2010} on the  de-convolved image, which results in accurate total flux measurements even for sources which are marginally resolved.
The flux measurement errors are estimated based on fitting errors of the 2D gaussian fits to the sources according to \citet{Condon1997}.
Data release 1 of A$^3$COSMOS contains continuum observations of 1544 ALMA pointings, covering a total of $\sim243\,\rm{arcmin}$.
We excluded all objects outside of the half primary beam width of each image.
Additionally, we only include images that have a resulting synthesised beam minor axis of $>0.5^{\prime\prime}$ in order to avoid that the sources are significantly resolved and fluxes are missed.

The datasets that we include have resulting synthesised beam sizes of $0.5^{\prime\prime} - 1.8^{\prime\prime}$ with median beam size of $1.1^{\prime\prime}$, and the median continuum sensitivity has an RMS of $126\,\rm{\mu Jy/beam}$ with standard deviation of a $91\,\rm{\mu Jy/beam}$, after applying primary beam corrections.

\begin{figure*}
	\includegraphics[width=\linewidth]{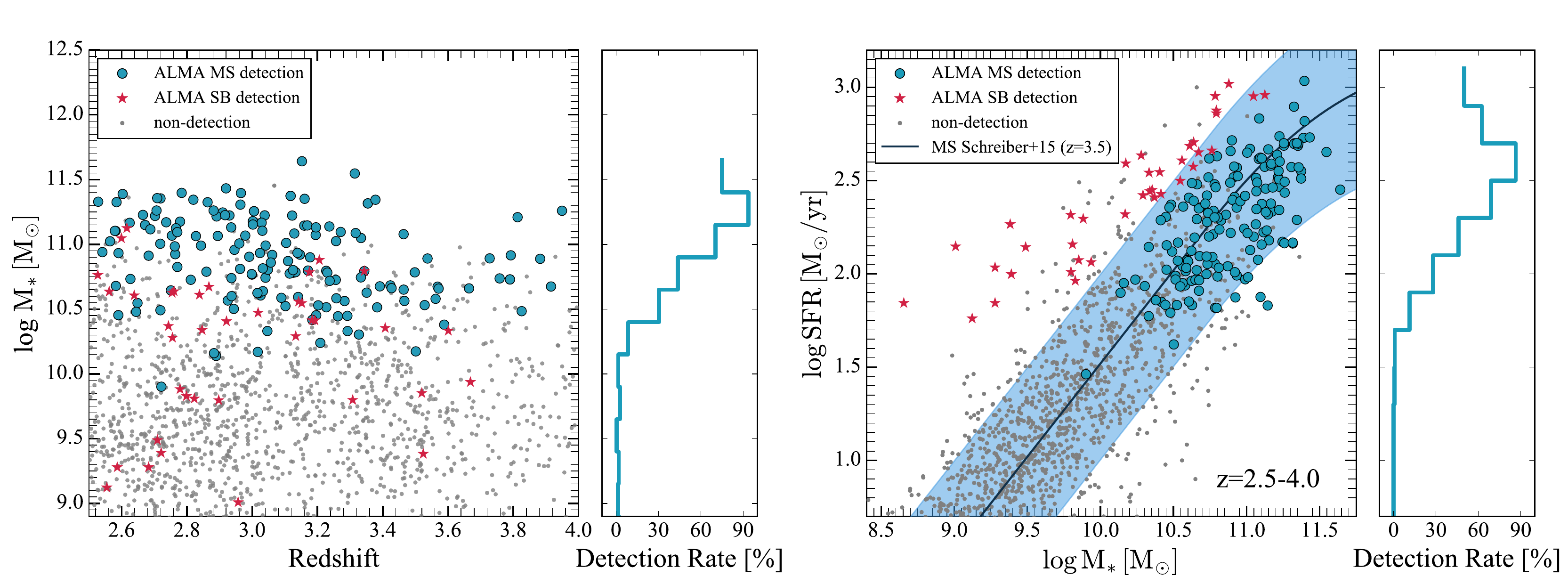}
    \caption{{\bf Left:} The distribution of our sample in  stellar mass ($M_{\ast}$) versus photometric redshift ($z_{\rm{ph}}$).
    {\bf Right:} The star formation rate (SFR) and stellar mass diagram showing the main-sequence of galaxies at $z=2.5-4.0$.
    The solid line represents the main-sequence at $z=3.5$ measured in \citet{Schreiber2015}, and the blue band indicates $\pm0.5\,\rm{dex}$ from the relation (see section \ref{sec:MS} for more details).
    For both panels, blue points and red stars indicate our main-sequence (MS) and starburst (SB) galaxies sample detected in ALMA band 6 or band 7
    ($>3\,\sigma$ signal at the prior position), respactively. Starbursts are defined as galaxies with SFRs $3\times$ above the main-sequence at their redshift. Gray dots show the sample that lies within the ALMA field of view (i.e. within half primary beam width), but did not yield a detection.
	For FIR detected galaxies, the total SFR is calculated from both $L_{\rm{UV}}$ and $L_{\rm{IR}}$ (see section \ref{sec:totalsfr}). For non-detected sources, SFRs are obtained from SED fitting to the optical to near infrared photometry in the COSMOS2015 catalogue.
    Most FIR detections are concentrated at the massive
    ($M_{\ast} \gtrsim 10^{10}\,\rm{M_{\odot}}$) end of the original catalogue.
    }
    \label{fig:zMass}
\end{figure*}

\subsection{Star Forming Galaxy Sample Selection}
Starting from the full A$^3$COSMOS dataset, we select all ALMA pointings that contain a galaxy at $z=2.5-4.0$, using the photometric redshifts from the COSMOS2015 catalogue. In particular, we only select galaxies with accurately determined photometric redshifts, $\delta_{\rm{z_{\rm{ph}}}}/(1.0+z_{\rm{ph}}) < 0.2$, where $\delta_{\rm{z}_{\rm{ph}}}$ corresponds to the $16\,\rm{th}$ to $84\,\rm{th}$ percentile width of the $z_{\rm{ph}}$ estimation in the COSMOS2015 catalogue.

Stellar masses are also taken from the COSMOS2015 catalogue, which have been estimated using the $\rm{LePhare}$ SED fitting code \citep{Arnouts2002,Ilbert2006} with  stellar population synthesis models of \citet{BC03} combining exponentially declining and delayed star formation histories (SFH). Solar and half solar metallicities are considered.
The median of the marginalised probability distribution function (PDF) is used throughout this paper ($\rm{mass\_med}$ column in the COSMOS2015 catalogue), and 16th, 84th percentile of the PDF is used as an estimate of the mass uncertainty.

Active galactic nuclei (AGNs) could potentially affect our analysis, as massive galaxies are frequently identified as hosts of AGNs.
Although the contribution of AGN emission is small in the FIR wavebands \citep[e.g.][]{Hatziminaoglou2010,Bethermin2015}, an AGN could outshine or significantly contribute to the rest-frame UV emission of its host galaxy, which would significantly affect the UV slope measurement.

To restrict our sample to star forming galaxies, we adopt the following additional selection criteria:

\noindent(1) The X-ray detected galaxies are identified as un-obscured AGNs.
Thus, we exclude galaxies that have counterparts in the deep Chandra X-ray catalogue of \citet{Civano2016}.
We cross-correlated our sample with the catalogue using a matching radius of $1.5^{\prime\prime}$. 46 of our sources have a clear X-ray detection and are thus excluded from further analysis.

\noindent(2)
We  further exclude potential AGNs through the rest-frame mid-infrared (MIR) photometry from Spitzer IRAC.
We employ MIR-AGN selection criteria developed by \citet{Donley2012},
which incorporate rest-frame mid-infrared colours to select obscured AGNs.
This method efficiently selects Compton-thick AGNs through their AGN heated dust colours in their rest-frame MIR wavebands. The MIR colour selection successfully avoids selecting moderate to high-redshift normal star forming galaxies at $z\sim0.5-3$ as AGNs.
This selection excludes only 44 sources from our sample.

Our final sample then includes 1512 galaxies at $z=2.5-4.0$ with median redshift $z_{\rm{ph}}=2.95$, and median stellar mass $\rm{log\,M_{\ast}}=9.82\,\rm{M_{\odot}}$.  172 galaxies out of this sample are detected at more than 3$\sigma$ with ALMA and 1338 only have upper limits. 




\section{Analysis}
\label{sec:analysis}

Among our final sample, the FIR detections are constrained to  massive ($M_{\ast}\gtrsim10^{10}\,\rm{M_{\odot}}$) galaxies only (Fig. \ref{fig:zMass}). Below this, most galaxies do not show significant dust continuum emission, down to the limits of current ALMA datasets in COSMOS. This is consistent with previous findings \citep[e.g.][]{Dunlop2017,Bouwens2016}. 
The mass distribution of our input catalogue peaks at $M_{\ast}\sim10^{9.5}\,\rm{M_{\odot}}$.
Fainter low mass galaxies with $M_{\ast} \lesssim 10^{9.0}\,\rm{M_{\odot}}$ are generally not included due to their larger photometric redshift uncertainties.

\subsection{Total SFR Estimation of ALMA Detected Sources}
\label{sec:totalsfr}

Interstellar dust absorbs UV emission from O-, B-type stars, and re-emits the energy at FIR wavelengths. A total measure of star formation rate ($\rm{SFR_{tot}}$) can be obtained by combining UV and IR luminosities of galaxies that are detected in FIR observations.
For the ALMA detected sample at $z_{\rm{ph}}=2.5-4.0$, we thus calculate star formation rates (SFRs) by employing the equations from \citet{Kennicutt1998} to the $L_{\rm{IR}}$ and $L_{\rm UV}$ estimated in section \ref{sec:lir} and \ref{sec:UVcalc}.
The SFRs from attenuation uncorrected UV luminosities ($\rm{SFR_{UV}}$) are calculated  as

\begin{equation}
	\rm{SFR_{UV}\,(M_{\ast}\,\rm{year^{-1}})} = 0.79\times10^{-28}\,L_{\nu_{1600}}\,(\rm{erg\,s^{-1}\,Hz^{-1}}),
\end{equation}
at $\lambda=1600\,$\AA, and the SFRs estimated from the IR luminosities ($\rm{SFR_{IR}}$) are calculated as

\begin{equation}
\rm{SFR_{IR}\,(M_{\ast}\,\rm{year^{-1}})} = 2.53\times10^{-44}\,L_{\rm{IR}}\,(\rm{erg\,s^{-1}}),
\end{equation}
where $L_{\rm{IR}}$ is the  FIR luminosity integrated over $\lambda=8-1000\,\rm{\mu m}$.
he total SFR ($\rm{SFR_{tot}}$) of the sample is then defined as $\rm{SFR_{tot}}=\rm{SFR_{UV}} + \rm{SFR_{IR}}$.

For ALMA non-detected galaxies, we directly use the SFR estimation from the COSMOS2015 catalogue as the total SFR of the galaxies, which is based on SED fitting employing a Calzetti extinction curve.

In the right panel of Fig. \ref{fig:zMass}, we show the $\rm{SFR-M_{\ast}}$ diagram of our sample. ALMA detected galaxies and non-detected galaxies together form a tight relation between stellar mass and SFR, which is known as the main-sequence of star forming galaxies \citep[e.g.][]{Elbaz2007,Daddi2007,Noeske2007,Peng2010}.

\subsection{$L_{\rm{IR}}$ Estimation}
\label{sec:lir}

The IR luminosities, $L_{\rm{IR}}$, of our galaxies are estimated from the ALMA continuum fluxes.
For this, we use the integrated flux on the cleaned images described in  section \ref{sec:data}.
$L_{\rm{IR}}$ is then estimated by scaling an SED template to the measured flux in ALMA band 6 $\lambda\sim1100-1400\,\rm{\mu m}$ or ALMA band 7 $\lambda\sim800-1100\,\rm{\mu m}$.
In particular, we use the empirical SED template that was previously derived for $z\sim3$ galaxies by \citet{AlvarezMarquez2016}.
This template is based on the SED library of \citet{Dale2014}, and is obtained by fitting stacked fluxes over a wide range of wavelengths ($100-1100\,\mathrm{\mu m}$). 
\citet{AlvarezMarquez2016} perform their stacking analyses for 22,000 galaxies selected by a Lyman break technique as a function of $L_{\rm{UV}}$, $\beta$, and $M_{\ast}$ after correcting the IR fluxes for clustering of galaxies and catalogue incompleteness.
From their results, we use, in particular, the SED template obtained in the mass range of $\rm{log}\,M_{\ast}/\mathrm{M_{\odot}}=10.25-10.75$, for which they find $\alpha_{\rm{dale}}=1.7$ in the \citet{Dale2014} description. As shown in appendix \ref{sec:SEDcompare}, this SED is also a good match to the average stacked {\it Herschel} FIR fluxes of our sample, and is thus an appropriate choice to estimate the total IR luminosities.

We redshift this SED template to the $z_{\mathrm{ph}}$ of each galaxy and rescale it to the measured ALMA flux. $L_{\rm{IR}}$ is then calculated by integrating the SED template over the rest-frame wavelength range $8-1000\,\rm{\mu m}$. Uncertainties on $L_{\rm{IR}}$ are estimated by propagating the flux measurement uncertainties.
For galaxies that are not detected in ALMA, we estimate the $3\,\sigma$ $L_{\rm{IR}}$ upper limit based on the image RMS, assuming the galaxies are not significantly resolved in the ALMA data, which have a resolution cut of $>0.5^{\prime\prime}$ (see Section \ref{sec:A3COSMOSsample}.

Clearly, the exact values of $L_{\rm{IR}}$ depend  on the assumed SED template, which we discuss in more detail in appendix \ref{sec:SEDcompare}. Using a wide range of templates, we find a systematic spread of $\pm0.2\,\rm{dex}$ on $L_{\rm{IR}}$, around the value derived with our baseline template. We therefore include this systematic uncertainty in addition to the measurement uncertainties when we report  $L_{\rm{IR}}$ values. Note, however, that the choice of a specific SED template as tested in the appendix does not affect our main conclusions.


\subsection{$L_{\rm{UV}}$ and $\beta$ Measurements}
\label{sec:UVcalc}

\begin{figure}
	\includegraphics[width=\columnwidth]{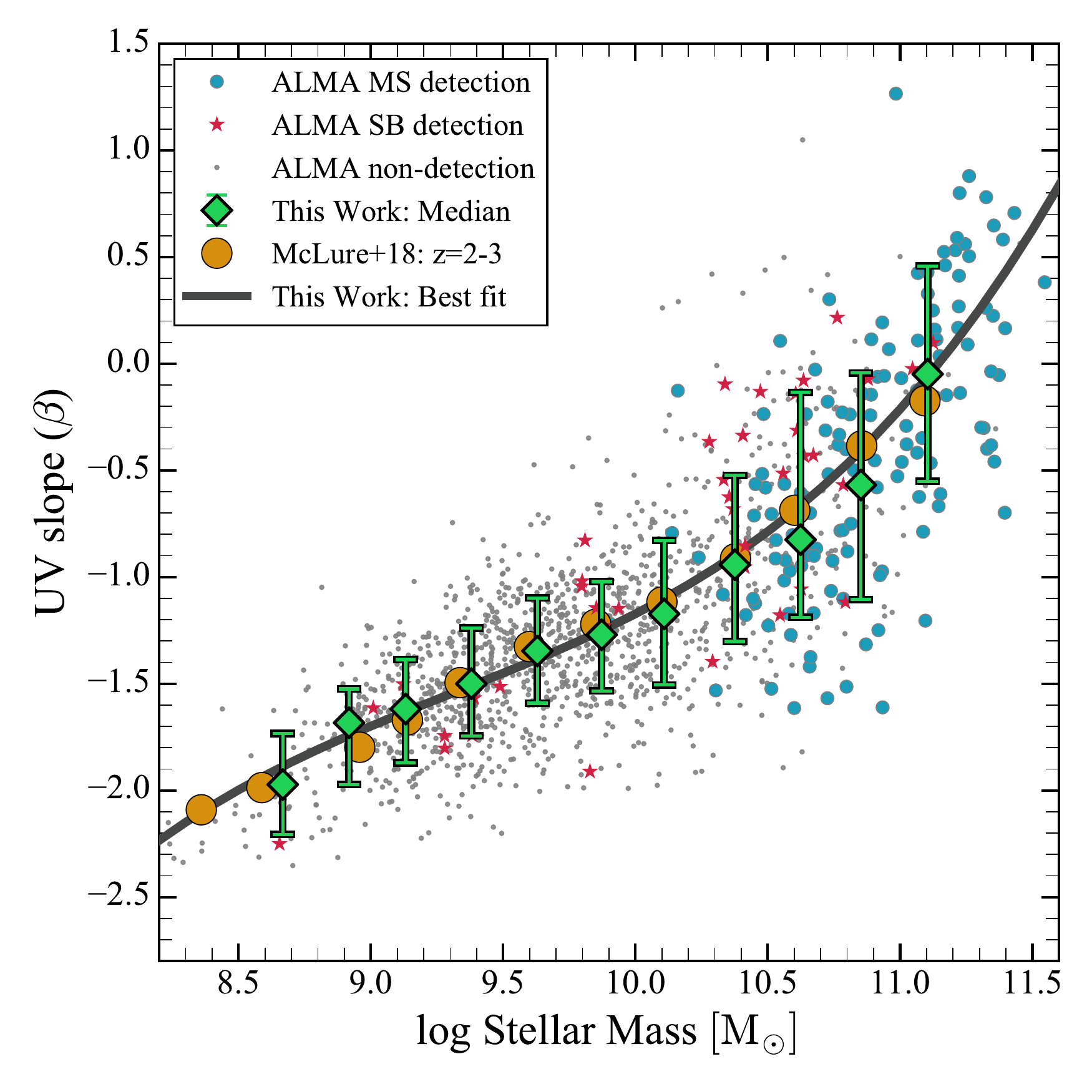}
    \caption{Observed UV slope ($\beta$) against stellar mass of our sample.
    Blue, red stars, and gray dots show individual galaxy measurement corresponding to ALMA detected main-sequence galaxies (MS), ALMA detected starbursts (SB) and non-detections, respectively. 
    The blue diamonds show median UV continuum slopes $\beta$, binned by stellar mass ($\rm{\Delta\,M_{\ast}}=0.25\,\rm{dex}$), and the error bars show the 16th and 84th percentiles indicating the distribution of $\beta$ in each bin.
    The black solid line shows our error weighted fit to the individual data points using a cubic polynomial function.
	Large yellow points show the results from \citet{Mclure2018}, who use a sample of star forming galaxies at $z=2-3$ observed in the HUDF, CANDELS GOODS-S, CANDELS UDS, and UVISTA surveys.
    \citet{Mclure2018} and our study give essentially identical results on the $\beta$-$\rm{M_{\ast}}$ relation, which assures us that our sample selected from a wider area dataset has a properly sampled $\beta$ distribution.
    }
    \label{fig:MassBeta}
\end{figure}

The UV spectral slopes ($\beta$) of our sample are estimated by employing SED fitting to photometric data from the COSMOS2015 catalogue, including broad-, intermediate-, and narrow-band filter observations at optical to NIR wavelengths after applying offsets and foreground extinction corrections  following the description in the Appendix A.2 of \citet{Laigle2016}. The photometry is then fitted using the code ZEBRA+ in combination with stellar synthesis templates based on \citet{BC03} of varying star formation histories and metallicities, to which we added emission lines and nebular continuum emission \citep[see][]{Oesch2010}.
To keep our results unbiased, we assumed two sets of extinction curves of local star forming galaxies \citep{Calzetti2001} and 
that measured for the SMC \citep{Gordon2003}. 

For each SED model, we calculate $\beta$ over the wavelength range of $1500<\lambda<2500\,$\AA, and we derive a marginalised probability distribution function of $\beta$ for each individual galaxy based on the $\chi^2$ values of each template. From this, we compute the median $\beta$ value and its uncertainties based on the 16th and 84th percentiles. 
Similarly, the monochromatic UV luminosities are calculated from the best fit SEDs at  $1600$\AA, with respective uncertainties, where we define $L_{\rm{UV}}=\nu_{1600} L_{\nu_{1600}}$.

Figure \ref{fig:MassBeta} shows the $\beta$ estimations against stellar mass.
We fit the relation with a third order polynomial function to the error weighted individual data points resulting in the following:

\begin{multline}
	\beta	=	-1.17\,(\pm0.01) + 0.63\,(\pm0.03)\,X\\
    + 0.22\,(\pm0.02)\,X^{2} + 0.11\,(\pm0.02)\,X^{3}
\end{multline}
where $X=\rm{log}\,(M_{\ast}\,\rm{/10^{10}\,M_{\odot}})$.
As shown in the Fig. \ref{fig:MassBeta}, this relation is in excellent agreement with \citet{Mclure2018}, who used a complete galaxy sample from all prime extragalactic legacy fields (HUDF, CANDELS, GOODS-S, CANDELS UDS, and the UVISTA survey).
This indicates that our sample, selected within ALMA pointings, is not biased in its UV slope distribution as a function of stellar mass.





\subsection{Classifying Galaxies as Main-Sequence Galaxies or Starbursts}
\label{sec:MS}


In order to study how starburst galaxies affect the IRX-$\beta$ relationship, we classified our sample either as starburst or main-sequence galaxies using our best estimate of their SFR (i.e., $\rm{SFR_{tot}}$ for the FIR continuum detected galaxies, or $\rm{SFR_{SED}}$ for ALMA non-detections).
The main-sequence (MS) relation as a function of redshift is taken from \citet{Schreiber2015}, and we defined galaxies as starbursts when they have SFR three times larger than the MS at their photometric redshift.
The FIR continuum detected sample as well as non-detected galaxies form a tight relation in the diagram following the main-sequence found in \citet{Schreiber2015}. 228 galaxies in our sample are classified as starbursts. In the following, we will discuss SBs and MS galaxies separately.

\subsection{Stacking Analysis}
\label{sec:stack}

Although only $\sim10\,\%$ of our full sample is detected with ALMA at $>3\,\sigma$ significance, the average ALMA fluxes including both detected and non-detected galaxies carry important information, which is why we perform a stacking analysis.
The main difficulty we have to overcome is that our dataset includes images from different projects with varying
beam sizes/orientations. Hence, an image stack is not straightforward. After testing different methods, we decided to opt for the simplest solution and to follow \citet{Mclure2018}, who derive average IRX measurements for different bins of galaxy masses and UV continuum slopes by stacking the ALMA aperture flux measurements. We compare these flux stacks to image stacks for a sub-sample of spectroscopically confirmed galaxies in section 4.2.1.


For our stacking method, we first need an ALMA flux measurement for each galaxy. Since the A$^3$COSMOS catalogue only provides fluxes for detections, we work with the ALMA continuum images directly, and derive our own flux measurements. 
We measure the ALMA flux of each galaxy within an aperture with a fixed radius of 2 arcsec.
A small aperture correction is then applied, dividing these fluxes by the dirty beam area within this aperture. The fixed aperture radius was chosen to best reproduce the total flux measurements of individually detected sources in the A3COSMOS catalogue.

For galaxies with $>3\,\sigma$ detections in the A3COSMOS catalogue, we perform these measurements on the dirty images, while for undetected sources, we perform these measurements on the clean images. This ensures that in both cases the signal corresponds to a dirty beam, and that sidelobes from bright sources within the same image do not affect our measurements. 
The weights for the stacks are based on the RMS values that are corrected by primary beam attenuation at the positions of objects.

The results of our stacking analyses are listed in Table \ref{tab:StackResults}, and will be discussed in the following sections.

\begin{table*}
\begin{center}
    \caption{Results of the Stacking Analysis}
	\label{tab:StackResults}
    \begin{tabular}{cccccccc}
    \hline\hline
    Range & \# of sources & <z> & <$\beta$> & log<$L_{\rm{UV}}$> & log<$M_{\ast}$> &  log$L_{\rm{IR}}$& logIRX\\
    & & & & [$\rm{L_{\odot}}$] & [$\rm{M_{\odot}}$] & [$\rm{L_{\odot}}$] & \\
   	\hline
    \multicolumn{8}{c}{Stack in bins of $\beta$ at $\rm{log}\,M_{\ast}/\rm{M_{\odot}}$ >10}\\
    \hline
    $-2.2 <\beta< -1.5$ & 26 & $3.31\pm0.32$ & $-1.58\pm0.11$ & $10.91\pm0.20$ & $10.13\pm0.25$ &  $11.84\pm0.22$ & $0.89\pm0.26$\\
    $-1.5 <\beta< -1.0$ & 121 & $3.19\pm0.34$ & $-1.20\pm0.14 $ & $10.75\pm0.30$ & $10.26\pm0.24$  & $11.44\pm0.23$ & $0.64\pm0.32$\\
    $-1.0 <\beta< -0.5$ & 131 & $3.14\pm0.36$ & $-0.82\pm0.13$ & $10.53\pm0.44$ & $10.44\pm0.29$ & $11.85\pm0.21$ & $1.40\pm0.27$\\
    $-0.5 <\beta< 0.0$ & 72 & $2.96\pm0.37$ & $-0.24\pm0.14$ & $10.18\pm0.29$ & $10.74\pm0.31$ & $12.11\pm0.21$ & $1.89\pm0.26$\\
    $0.0 <\beta< 0.5$ & 27 & $2.93\pm0.27$ & $0.25\pm0.13$ & $9.95\pm0.19$ & $11.09\pm0.32$ & $12.42\pm0.21$ & $2.46\pm0.27$\\
    $0.5 <\beta< 1.0$ & 12 & $2.90\pm0.09$ & $0.68\pm0.16$ & $9.93\pm0.19$ & $11.25\pm0.01$ & $12.51\pm0.21$ & $2.70\pm0.28$\\
    \hline
    \multicolumn{8}{c}{Stack in bins of $\beta$ at $\rm{log}\,M_{\ast}/\rm{M_{\odot}}=9-10$}\\
    \hline
    $-2.2<\beta< -1.5$ & 239 & $3.12\pm0.40$ & $-1.67\pm0.15$ & $10.30\pm0.28$ & $9.34\pm0.25$ & $<10.68$ & $<0.58$\\
    $-1.5 <\beta$ & 353 & $3.01\pm0.36$ & $-1.29\pm0.19$ & $10.33\pm0.26$ & $9.63\pm0.23$ & $<10.51$ & $<0.28$\\
    \hline
    \multicolumn{8}{c}{Stack in bins of $\rm{log\,M_{\ast}}$}\\
   	\hline
    $11.0 <M_{\ast}< 12.0$ & 51 & $3.20\pm0.17$ & $-0.06\pm0.65$ & $10.26\pm0.34$ & $11.15\pm0.17$  & $12.45\pm0.21$ & $2.30\pm0.23$\\
    $10.5 <M_{\ast}< 11.0$ & 132 & $3.28\pm0.26$ & $-0.84\pm0.45$ & $10.64\pm0.38$ & $10.68\pm0.13$ & $11.98\pm0.22$ & $1.33\pm0.23$\\
    $10.0 <M_{\ast}< 10.5$ & 212 & $3.34\pm0.25$ & $-1.10\pm0.34$ & $10.71\pm0.32$& $10.23\pm0.15$ & $11.36\pm0.26$ & $0.75\pm0.29$\\
    $9.5 <M_{\ast}< 10.0$ & 318 & $3.40\pm0.26$ & $-1.36\pm0.24$ & $10.46\pm0.24$ & $9.72\pm0.14$  & $<10.53$ & $<0.19$\\
    \hline
    \multicolumn{8}{c}{Stack of Spec-z Sample}\\
    \hline
    $9.5 <M_{\ast}< 10.0$ & 61 & $3.44 \pm 0.41$ & $-1.23 \pm 0.26$ & $10.94 \pm 0.22$ & $10.08 \pm 0.20$ & $11.13\pm0.26$ & $0.24\pm0.27$ \\
    \hline
    \end{tabular}
	\end{center}
    \begin{flushleft}
	    $\ast$ The parentheses <> represent median values of the stacked sample, and the uncertainty correspond to standard deviations.
	\end{flushleft}
\end{table*}

\section{Results}

\subsection{The IRX-$\beta$ Relation at $z=2.5-4.0$}
In the following sections, we first analyse the IRX-$\beta$ relation by separating our sample at a threshold stellar mass above and below $M_{\ast}=10^{10}\,\rm{M_{\odot}}$.

\subsubsection{IRX-$\beta$ Relation of Massive Galaxies ($M_{\ast} > 10^{10}\,M_{\odot}$)}
\begin{figure*}
	\includegraphics[width=18cm]{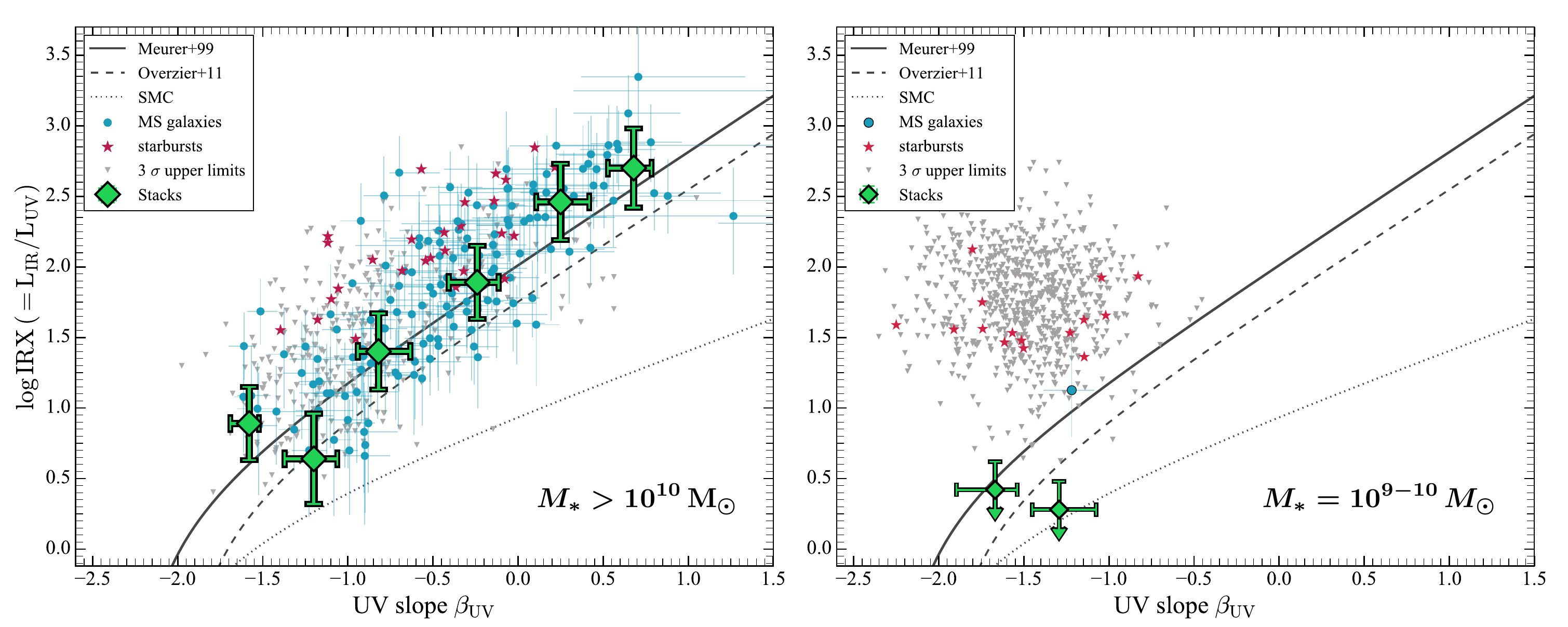}
    \caption{ 
    IRX-$\beta$ diagram of massive ($\rm{M_{\ast} > 10^{10}\,M_{\odot}}$, {\bf left}) and less massive ($10^{9}\,M_{\odot} < \rm{M_{\ast} < 10^{10}\,M_{\odot}}$, {\bf right}) galaxies at $z\sim2.5-4.0$. 
    Blue points are individual FIR continuum detections of main-sequence galaxies (with
    significance $>3\,\sigma$). Red stars indicate the FIR continuum detected starburst galaxies (see section \ref{sec:MS}).  Grey triangles show 
    $3\sigma$ upper limits for non-detected sources.
    Large green diamonds are results from a stacking analysis of the main-sequence sample. The stacks are performed in bins of $\beta$ ($-2.2<\beta < -1.5$, $-1.5<\beta < -1.0$, $-1.0<\beta < -0.5$, $-0.5<\beta < 0.0$, $0.0<\beta < 0.5$, $0.5<\beta < 1.0$, $1.0<\beta < 1.5$).
    Three relations are also shown: the relation of local starburst galaxies \citep[solid line;][]{Meurer1999}, the updated relation for local galaxies with wider apertures for UV photometry \citep[dashed line;][]{Overzier2011}, and an SMC-like dust attenuation relation \citep[dotted line, e.g.][]{Prevot1984}.
    At the high mass end ($\rm{M_{\ast} > 10^{10}\,M_{\odot}}$), individual detections and the stacks show that the massive galaxies on the main sequence at this redshift range are consistent with the M99 relation. Starburst galaxies are generally offset and have $\sim0.5\,\rm{dex}$ larger IRX values than  main sequence galaxies.
    Less massive galaxies ($10^{9}\,M_{\odot} < \rm{M_{\ast} < 10^{10}\,M_{\odot}}$) are mostly non-detected, and stacks tentatively show lower values in IRX than the M99 relation.
    }
    \label{fig:IRXbeta_highMass}
\end{figure*}

The left panel of Figure \ref{fig:IRXbeta_highMass} shows the IRX-$\beta$ relation of the massive galaxies.
Our sample covers a wide range of $\beta\sim-1.5$ to $1.0$, which is very close to the range originally sampled by M99.
Individually detected galaxies show a fairly tight IRX-$\beta$ relation,  consistent with that of local starburst galaxies. 

In order to compute the average IRX of the whole population (not only of ALMA detections), we performed a stacking analysis on the full main-sequence galaxy sample, as described in section \ref{sec:stack}.
The mean stacked IRX measurements in bins of $\beta$ almost perfectly follow the local relation relation of M99 (Fig. \ref{fig:IRXbeta_highMass}).
This is consistent with our pilot study \citep{Fudamoto2018}, and with several previous studies employing stacking analyses \citep[e.g.][]{AlvarezMarquez2016,Koprowski2018}.

Our $L_{\rm{IR}}$ estimation could vary by up to $\sim0.2\,\rm{dex}$ when applying different SED models  (see Appendix). Given this uncertainty, we cannot conclusively distinguish between the original M99 relation or its updated version \citep{Overzier2011} for our massive galaxy sample. However, we can clearly rule out the SMC relation for these galaxies, as the systematic uncertainties due to the SEDs are much smaller than the  difference between the SMC and the M99 relation.

The observed IRX-$\beta$ relation is relatively tight. The random scatter around the mean relation is only $\sigma_\mathrm{IRX}\sim 0.27\,\rm{dex}$ (excluding the systematic uncertainty due to the IR SED). 
This value is significantly smaller than the $\sim 0.5\,\rm{dex}$ reported in \citet[][]{Fudamoto2018}. The much larger statistics clearly improved our measurement. Additionally, our previous UV continuum slopes were derived from power-law fits to the photometry rather than using properly sampled UV slopes probability distributions from SED fitting. 


Individually, most of the $3\,\sigma$ upper limits from the non-detected, massive galaxies are consistent with  random scatter.
Given our detection limits, only 3 galaxies lie more than $3\times\sigma_\mathrm{IRX}$ below the M99 relation.
From UV observations, these galaxies do not exhibit any different features (such as errors in $\beta$ and/or $L_{\rm{UV}}$) compared to other galaxies.
This indicates that the diversity of dust attenuation properties in massive galaxies is rather limited. Nevertheless, massive galaxies with attenuation curves similar to the SMC may still exist. Deeper ALMA data on currently undetected sources will be required to test this.


Our results confirm that for high-redshift, massive galaxies, on average, there is no noticeable evolution of the IRX-$\beta$ relationship since $z\sim4.0$ to the present Universe, as already indicated by several studies \citep[e.g.][]{Bouwens2016,Koprowski2018,Mclure2018}

Starburst galaxies (as identified in section \ref{sec:MS}) generally have large IRX values,  on average  $\rm{\sim 0.5\,dex}$ above the local relation.
This can be expected, as heavily obscured regions together with a small fraction of non-obscured regions ("holes in dust shields") easily deviate from the M99 relation, because their UV and IR fluxes no longer come from same region of a galaxy \citep[see e.g.][]{Popping2017,Narayanan18a}.

\subsubsection{IRX-$\beta$ Relation of Lower Mass Galaxies ($M_{\ast} < 10^{10}\,M_{\odot}$)}

For our lower mass sample, we have 970 flux measurements, but only 16 galaxies are detected at $>3\,\sigma$. Except for one source, the detected galaxies are all classified as starburst (see Section \ref{sec:MS}).
Out of the 970 galaxies, only 80 were the  main target of the ALMA observations (typical sensitivity $0.08\,\rm{mJy/beam}$) and the rest only have serendipitous coverage within the primary beam of a neighbouring target. As such, most of the observations were not deep enough to expect an individual detection, in particular, since the lower mass sources ($\rm{log\,M_{\ast}/M_{\odot}\sim 9.0-10.0}$) on the main-sequence are mostly dominated by blue UV colours ($\beta\lesssim-1.0$), in contrast to their massive counterparts. This trend is seen in Section \ref{sec:UVcalc}, from the $\beta-M_{\ast}$ relationship (see also Fig. \ref{fig:MassBeta}), which already indicates that lower mass galaxies have a significantly lower dust content than higher mass ones.

As the low-mass galaxies are mostly non-detected, we have to rely on a stacking analysis to provide constraints on the IRX-$\beta$ relation. However, even when splitting the sample in only two bins of UV slope, we do not find a significant detection in the mean stacked fluxes (see Fig. \ref{fig:IRXbeta_highMass}).

For the redder sample, the $3\sigma$ upper limit of the stacked flux lies below the M99 relation and is more consistent with SMC dust extinction curve. However, we cannot conclusively rule out the M99 relation, given the IR SED uncertainties.
Nevertheless, our result is consistent with the study using deep ALMA band 6 continuum in the Hubble Ultra Deep Field \citep{Bouwens2016}, showing that the stacks of low mass galaxies ($M_{\ast} < 10^{9.75}\,\rm{M_{\odot}}$) typically lie below the IRX-$\beta$ relation inferred from an SMC-like dust extinction curve. Deeper observations of low mass galaxies at these redshifts will be required to confirm this.

Theoretically, this mass dependent difference is predicted from some  cosmological hydrodynamical simulations. In particular, \citet{Mancini2015} find that their simulated lower mass galaxies at $z>5$ follow an SMC-like dust extinction, while massive galaxies follow the local relation of M99. Our observations indicate that this trend may already be in place at $z\sim3-4$.

\subsection{IRX-$M_{\ast}$ Relation}
\label{sec:IRXmass}

\begin{figure*}
	\includegraphics[width=17cm]{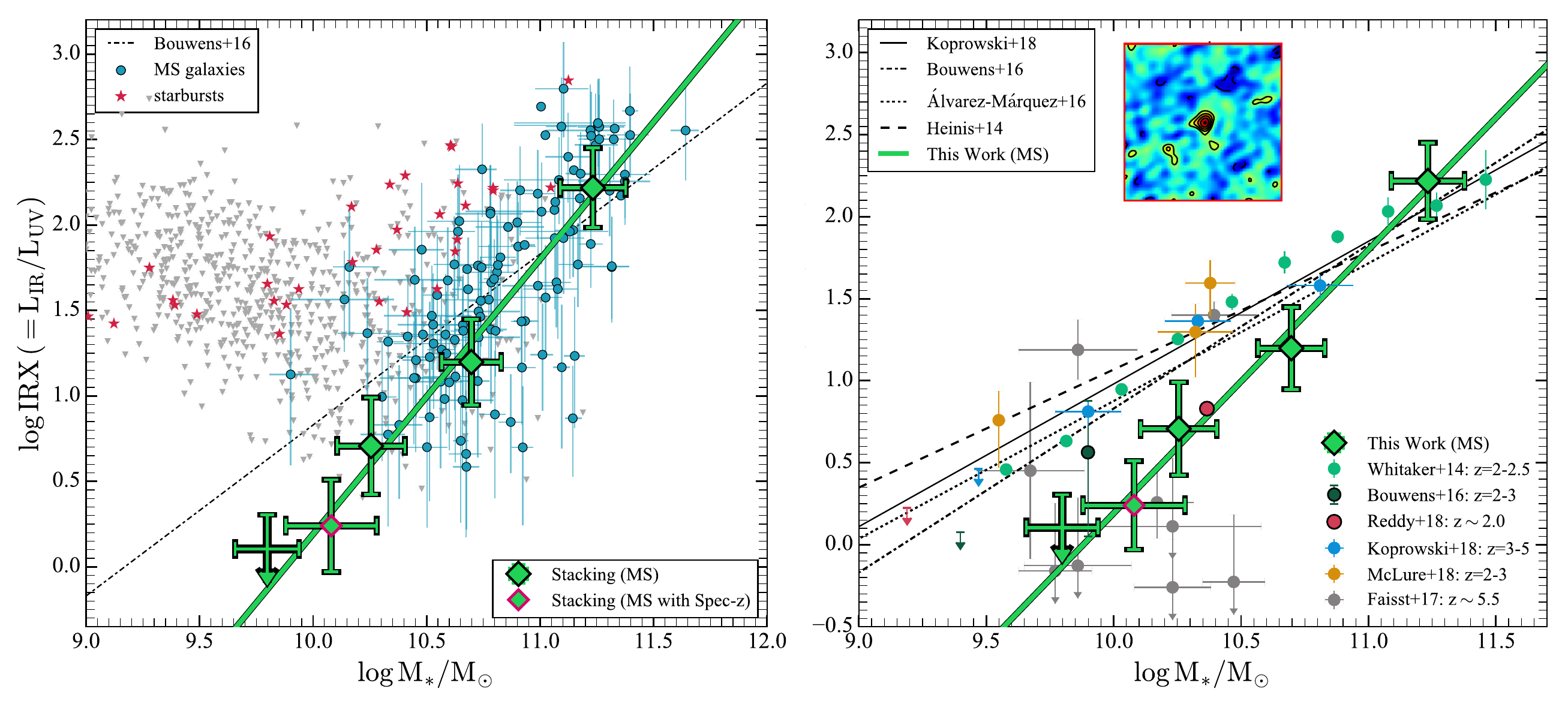}
    \caption{The IRX-$M_{\ast}$ diagram of  star forming galaxies at $z=2.5-4.0$. As in the previous plots, the blue points, red stars and grey points correspond to main-sequence galaxy detections, starbursts, and $3\sigma$ upper limits for non-detections, respectively.
    Dark green diamonds show our stacking results in bins of mass.
    The one green diamond with a red edge shows the stacking result for the spectroscopic redshift sample described in section \ref{sec:IRXpecz}.
    The dark green solid line shows our fitting result to the three mass bins at $M_{\ast}>10^{10}\,\rm{M_{\odot}}$. Our measurements reveal a tight correlation between IRX and $M_{\ast}$, with a very steep slope.
    In the right panel, we compare our results with previous studies at similar redshifts. Interestingly, both our individual detections and the stacks follow a significantly steeper slope of the IRX-$M_{\ast}$ relation than previously reported in the literature \citep[e.g.][]{Heinis2014,AlvarezMarquez2016,Bouwens2016,Koprowski2018}.
    In particular, our results imply significantly less dust extinction in low mass galaxies than inferred from previous relations.
    }
    \label{fig:IRXmass}
\end{figure*}

In the left panel of Figure \ref{fig:IRXmass}, we show the IRX-$M_{\ast}$ diagram of our sample.
The individual ALMA detections among our main-sequence sample show a relatively tight correlation between IRX and  $M_{\ast}$. As expected, starburst galaxies are located above the relation of main-sequence galaxies, indicating that they are more dust extinct at a fixed stellar mass.

In general, the individually detected MS galaxies are in good agreement with the ``consensus relationship" derived in \citet{Bouwens2016} based on several previous studies at $z>2$. However, we see an indication for a steeper IRX-$M_{\ast}$ relation, as the majority of the most massive galaxies lie above the previous relation, while intermediate mass galaxies ($M_{\ast}\sim10^{10.5}\,\rm{M_{\odot}}$) seem to have lower IRX than the ``consensus relationship" and previous works \citep[e.g.][]{Bouwens2016,Mclure2018,Heinis2014,Whitaker2014,Bouwens2016,AlvarezMarquez2016}. 

The finding of a significantly steeper IRX-$M_{\ast}$ becomes very clear when stacking our full galaxy sample in bins of stellar mass, including non-detections. 
The results of our stacking analysis at $M_{\ast}>10^{9.5}\,\rm{M_{\odot}}$ are listed in Table \ref{tab:StackResults} and shown in Figure \ref{fig:IRXmass}.
In particular, at $M_{\ast}<10^{10}\,\rm{M_{\odot}}$, even our stack of more than 300 galaxies did not reveal a detection, indicating a significantly lower IRX value than expected from previous relations, by $\gtrsim0.5\,\rm{dex}$.




The linear fit to the detected stacked data points (the first three green diamonds) results in the following equation  for the IRX-$M_{\ast}$ relation:

\begin{equation}
	\rm{log\,IRX} = (1.64 \pm 0.13)\times \rm{log\,}M_{\ast} - (16.18\pm 1.35 \pm0.2),
    \label{eqn:IRXmass}
\end{equation}
where, $M_{\ast}$ is in  units of $M_{\odot}$, and the $\pm0.2\,\rm{dex}$ uncertainty comes from the IR SED shape. This fit is also shown in Figure \ref{fig:IRXmass}.


\subsubsection{Confirmation from a Spectroscopic Redshift Sample at $z\sim3.5$}
\label{sec:IRXpecz}

In order to examine if our IRX-$\beta$ relation might be affected by photometric redshift uncertainties or whether our aperture stacking method might be biased, we have used an alternative stacking method on a sample with redshifts confirmed through spectroscopic observations.
The program we used is $\rm{2015.1.00379.S}$ (PI: E. Schinnerer). The sample has redshift measurements from the VUDS spectroscopic survey \citep{LeFevre2015}, and is observed with ALMA band 6 (at $\nu_{\rm{obs}}\sim 240\,\rm{GHz}$).
The sample includes $61$ galaxies ranging in stellar mass between $10^{9.5}\,\rm{M_{\odot}}<M_{\ast}<10^{10.5}\,\rm{M_{\odot}}$ (median $M_{\ast}$ of $10^{10.1}\,\rm{M_{\odot}}$) with  spectroscopic redshifts ranging from 3.0 to 4.74 (median redshift of $z=3.44$).
Using this sample, we employed a stacking method in the visibility domain, as described in \citet{Lindroos2015}.

The stacked image shows a $6.1\,\sigma$ detection in the FIR continuum emission, with a flux density of $56\pm9\,\rm{\mu Jy/beam}$, consistent with the value  obtained by our stack employing aperture flux averages. The stacked image is shown as an inset in the right panel of Figure \ref{fig:IRXmass}.
The UV slope, IR luminosity, and UV luminosity, are calculated using the same method as described in section \ref{sec:analysis}.

This stacking result is perfectly consistent with the IRX-$M_{\ast}$ relation inferred from individually detected galaxies and from stacks of our entire sample. This confirms that our finding of a steep IRX-Mass relation is robust and is neither due to photometric redshift uncertainties nor due to our stacking method.

\subsubsection{Comparison with IRX-$\rm{M_{\ast}}$ Relations From Previous Analyses}

In the right panel of Figure \ref{fig:IRXmass}, we compare our findings with several results in the literature.
These include stacking analyses of single dish sub-mm telescope data, {\it Spitzer} MIPS, and/or {\it Herschel} SPIRE data using rest-frame UV/Optical galaxy selections \citep{AlvarezMarquez2016,Koprowski2018,Heinis2014,Whitaker2017,Reddy2018}.
Additionally, we show the stacking results of ALMA data \citep{Bouwens2016,Mclure2018}, which utilise a smaller field of view, but have a higher angular resolution than the previous observations.

A few of the previous studies are consistent with our finding of a very steep IRX-$M_{\ast}$ relation \citep{Bouwens2016,Reddy2018}. However, most studies report a shallower slope, such that they find higher IRX values for lower mass galaxies (i.e. $\rm{log} M_{\ast} < 10.5$) \citep{Heinis2014,Whitaker2014,AlvarezMarquez2016,Mclure2018,Koprowski2018}.

Interestingly, our result of a lower IRX for $\log M_\ast/M_\odot \simeq 10$ galaxies is consistent with the few measurements that currently exist for $z\sim5-6$ galaxies \citep[][]{Capak2015,Faisst2017}. In fact, many of the galaxies at those redshifts show $\log$IRX$<0$ and are currently undetected in ALMA. Together with our results, this may indicate a possible evolution of the dust properties at $z>2$, before the peak of the cosmic star formation history. We will analyse this in detail in a future paper.

\subsubsection{Potential Biases of IRX Measurements}
\label{sec:caveats}
The origin for the differences with respect to previous measurements is not immediately clear. However, there are several assumptions and biases that can possibly affect both previous studies as well as ours. The most obvious ones, which we discuss in the following are (1) the dust continuum SED shape and (2) potential measurement biases in stacking of low resolution data (e.g. Herschel).

\textit{(1) Biases due to Different FIR SEDs: }
The assumed shape of the FIR SED  clearly affects the IR luminosity estimates, in particular, if the SED shape changes with stellar mass or luminosity.
Our study relies on mm/submm observations of dust continuum emission, which is in the Rayleigh-Jeans regime of the FIR SED from high-redshift galaxies.
Thus, mm/submm observations could be biased toward emission from low temperature dust. If the dominant part of the FIR emission of $z\sim3$ galaxies arises from higher temperature dust, our $L_{\rm{IR}}$ (thus IRX) estimation of individual or stacked data could be systematically lower than the intrinsic distribution of $L_{\rm{IR}}$ and IRX at $z\sim3$.

As seen in the Figure \ref{fig:IRXmass}, while our IRX measurements agree very well with previous studies for high mass galaxies, we find a decrement of $\sim0.5\,\rm{dex}$ for lower mass sources. Most of the previous analyses rely on stacking of shorter wavelength data than what we have access to here, i.e. they can potentially probe wavelengths closer to the peak of the SED.
While our assumed template SED was derived through one of the stacks \citep{AlvarezMarquez2016} and it is consistent with dust temperature measurements of massive galaxies at these redshifts \citep[$T_d=40\,\rm{K}$ ; e.g.][]{Fudamoto2018,Schreiber2018}, it is of course possible that the dust temperature and thus the SED shape changes systematically with galaxy mass. 
In order to flatten our IRX-mass relation, a higher dust temperature would be required for lower mass  or lower IR luminous galaxies (i.e. $T_{\rm{d}}>50\,\rm{K}$ in $\rm{log}\,(M_{\ast}/M_{\odot})<10.5$ galaxies). However, no significant evidence for such trends currently exists from {\it Herschel} stacks at these redshifts \citep{Schreiber2018}, and lower redshift galaxies seem to show the opposite trend  \citep[e.g.][]{Casey2018b}.

The variation of dust temperature also affects dust mass estimations, as the inferred dust mass is inversely proportional to the temperature \citep[e.g.][]{Scoville2016}. Therefore, an uncertain dust temperature variation results in an additional systematic uncertainty in the molecular gas mass estimation based on dust masses \citep[][]{Scoville2016,Schinnerer2016}.
It is clear that a direct study of the dust temperature of high-redshift galaxies is crucial to fully understand their FIR emission and to resolve these discrepancies.

\textit{(2) Clustering Biases in Stacking Analyses: }
In contrast to our relatively high-resolution ALMA measurements of individual sources, previous studies have mostly relied on stacking of very low resolution data, which can be significantly affected by clustering biases \citep[see][]{Magnelli2014,Bethermin2015}. When attempting to stack the Herschel fluxes of our galaxies (see Appendix), it became clear that this clustering bias can become large and difficult to correct for, in particular for low mass, low luminosity galaxies. It therefore seems very difficult to make further progress in resolving this issue with current data.

\section{Conclusions}
We have analysed the IRX-$\beta$/$M_{\ast}$ dust attenuation relations of star forming galaxies at $z\sim2.5-4.0$ using all the publicly available ALMA archival data in the COSMOS field. In particular, we include all primary targeted galaxies as well as serendipitously covered sources within the primary beam width of existing ALMA continuum images that were processed by the A$^3$COSMOS project team \citep{Liu2018}.
The ALMA data were cross-matched with the COSMOS2015 optical-NIR photometric catalogue \citep{Laigle2016}.
Our final sample consists of 1510 ALMA flux measurements with beam size $0.5-1.8\,\rm{arcsec}$ and a typical sensitivity of $126\,\rm{\mu Jy/beam}$. In total only $\sim10\%$ of the $z\sim2.5-4.0$ sample is detected with signal-to-noise ratio $>3$. The detection rate is highly mass dependent and decreases significantly below a stellar mass $\log (M_\ast/M_\odot)
\sim 10.5$ (Fig \ref{fig:zMass}).

The analysis of these data leads to the following conclusions:

({\romannumeral 1}) For massive galaxies with stellar mass greater than $10^{10}\,M_{\odot}$, the IRX-$\beta$ relation is in very good agreement with the local relation derived by \citet{Meurer1999} over a wide range of the UV slope $\beta$ ($\beta\sim -1.7$ to 1.0).
The scatter around the mean relation is small, only $0.27\,\rm{dex}$.
This indicates that the dust properties do not evolve significantly for massive galaxies between $z\sim4.0$ and the present day Universe.
This trend is also consistent with the theoretical prediction of hydrodynamic simulations in previous works \citep[e.g.][]{Mancini2015}. 

({\romannumeral 2}) The detection rate drastically decreases in less massive galaxies, and individual objects and stacking analyses only provide upper limits on the flux measurements. For galaxies with $M_{\ast}<10^{10}\,\rm{M_{\odot}}$ upper limits on their average IRX values fall below the local relation and are more consistent with an SMC dust attenuation relation. 
This is consistent with previous, deep ALMA surveys \citep[e.g.][]{Bouwens2016}.
However, some tension exists with other works in the literature. For instance, \citet{Mclure2018} find that galaxies with even lower mass ($M_{\ast}\sim10^{9.5}\,\rm{M_{\odot}}$) at these redshifts still follow the local IRX-$\beta$ relation quite well.
Deeper data on a larger sample of low mass galaxies are needed to distinguish between these two findings.

({\romannumeral 3}) Our sample shows a tight correlation between IRX and galaxy stellar mass. The slope is found to be steeper than reported in previous analyses which were mostly based on stacks of lower angular resolution data (e.g. Herschel) or based on smaller samples. This steep slope implies that $z\sim3-4$ galaxies with masses below $M_{\ast}<10^{10}\,\rm{M_{\odot}}$ suffer 3$\times$ less dust extinction at a given UV luminosity than previous results might have implied. These low extinction values are consistent with early results from small sample of $z\sim5.5$ galaxies and they could be a first hint at a possible evolution of the dust extinction properties in the early Universe, at $z>2$, compared to later cosmic times.


({\romannumeral 4})
Our results highlight the need of detailed studies on the FIR continuum emission of high-redshift galaxies.
In particular, our analysis is based only on the mm/submm ALMA fluxes, from which we derive an estimate of the galaxies' infrared luminosities. Systematic changes of the dust temperature or SED shapes as a function of stellar mass could affect the slope of our inferred IRX-mass relation. To address these possible systematic biases in the future, the FIR SEDs of high-redshift galaxies will need to be studied carefully through direct observations of individual sources at higher frequency (e.g. ALMA band 8/9/10), or through indirect inferences from their local analogues.

\section*{Acknowledgements}

This paper makes use of the following ALMA data: 
\path{ADS/JAO.ALMA#2011.0.00064.S}, 
\path{ADS/JAO.ALMA#2011.0.00097.S}, 
\path{ADS/JAO.ALMA#2011.0.00539.S}, 
\path{ADS/JAO.ALMA#2011.0.00742.S}, 
\path{ADS/JAO.ALMA#2012.1.00076.S}, 
\path{ADS/JAO.ALMA#2012.1.00323.S}, 
\path{ADS/JAO.ALMA#2012.1.00523.S}, 
\path{ADS/JAO.ALMA#2012.1.00536.S}, 
\path{ADS/JAO.ALMA#2012.1.00919.S}, 
\path{ADS/JAO.ALMA#2012.1.00952.S}, 
\path{ADS/JAO.ALMA#2012.1.00978.S}, 
\path{ADS/JAO.ALMA#2013.1.00034.S}, 
\path{ADS/JAO.ALMA#2013.1.00092.S}, 
\path{ADS/JAO.ALMA#2013.1.00118.S}, 
\path{ADS/JAO.ALMA#2013.1.00151.S}, 
\path{ADS/JAO.ALMA#2013.1.00171.S}, 
\path{ADS/JAO.ALMA#2013.1.00208.S}, 
\path{ADS/JAO.ALMA#2013.1.00276.S}, 
\path{ADS/JAO.ALMA#2013.1.00668.S}, 
\path{ADS/JAO.ALMA#2013.1.00815.S}, 
\path{ADS/JAO.ALMA#2013.1.00884.S}, 
\path{ADS/JAO.ALMA#2013.1.00914.S}, 
\path{ADS/JAO.ALMA#2013.1.01258.S}, 
\path{ADS/JAO.ALMA#2013.1.01292.S}, 
\path{ADS/JAO.ALMA#2015.1.00026.S}, 
\path{ADS/JAO.ALMA#2015.1.00055.S}, 
\path{ADS/JAO.ALMA#2015.1.00122.S}, 
\path{ADS/JAO.ALMA#2015.1.00137.S}, 
\path{ADS/JAO.ALMA#2015.1.00260.S}, 
\path{ADS/JAO.ALMA#2015.1.00299.S}, 
\path{ADS/JAO.ALMA#2015.1.00379.S}, 
\path{ADS/JAO.ALMA#2015.1.00388.S}, 
\path{ADS/JAO.ALMA#2015.1.00540.S}, 
\path{ADS/JAO.ALMA#2015.1.00568.S}, 
\path{ADS/JAO.ALMA#2015.1.00664.S}, 
\path{ADS/JAO.ALMA#2015.1.00704.S}, 
\path{ADS/JAO.ALMA#2015.1.00853.S}, 
\path{ADS/JAO.ALMA#2015.1.00861.S}, 
\path{ADS/JAO.ALMA#2015.1.00862.S}, 
\path{ADS/JAO.ALMA#2015.1.00928.S}, 
\path{ADS/JAO.ALMA#2015.1.01074.S}, 
\path{ADS/JAO.ALMA#2015.1.01105.S}, 
\path{ADS/JAO.ALMA#2015.1.01111.S}, 
\path{ADS/JAO.ALMA#2015.1.01171.S}, 
\path{ADS/JAO.ALMA#2015.1.01212.S}, 
\path{ADS/JAO.ALMA#2015.1.01495.S}, 
\path{ADS/JAO.ALMA#2015.1.01590.S}, 
\path{ADS/JAO.ALMA#2015.A.00026.S}, 
\path{ADS/JAO.ALMA#2016.1.00478.S}, 
\path{ADS/JAO.ALMA#2016.1.00624.S}, 
\path{ADS/JAO.ALMA#2016.1.00735.S}. 

ALMA is a partnership of ESO (representing
its member states), NSF (USA) and NINS (Japan), together
with NRC (Canada) and NSC and ASIAA (Taiwan) and KASI
(Republic of Korea), in cooperation with the Republic of Chile.
The Joint ALMA Observatory is operated by ESO, AUI/NRAO
and NAOJ.
This work was supported by the Swiss
National Science Foundation through the SNSF Professorship grant
157567 'Galaxy Build-up at Cosmic Dawn'.
PL, DL, and ES acknowledge funding from the European Research Council (ERC) under the European 
Union's Horizon 2020 research and innovation programme (grant agreement No. 694343).
SL acknowledge funding from SCHI 536/9-1.
EFJA acknowledge support of the Collaborative Research Center 956, subproject A1, funded by the Deutsche Forschungsgemeinschaft (DFG).  




\bibliographystyle{mnras}
\bibliography{./ref.bib}




\appendix

\section{Comparison of SED Templates}
\label{sec:SEDcompare}

\begin{figure}
	\includegraphics[width=8.5cm]{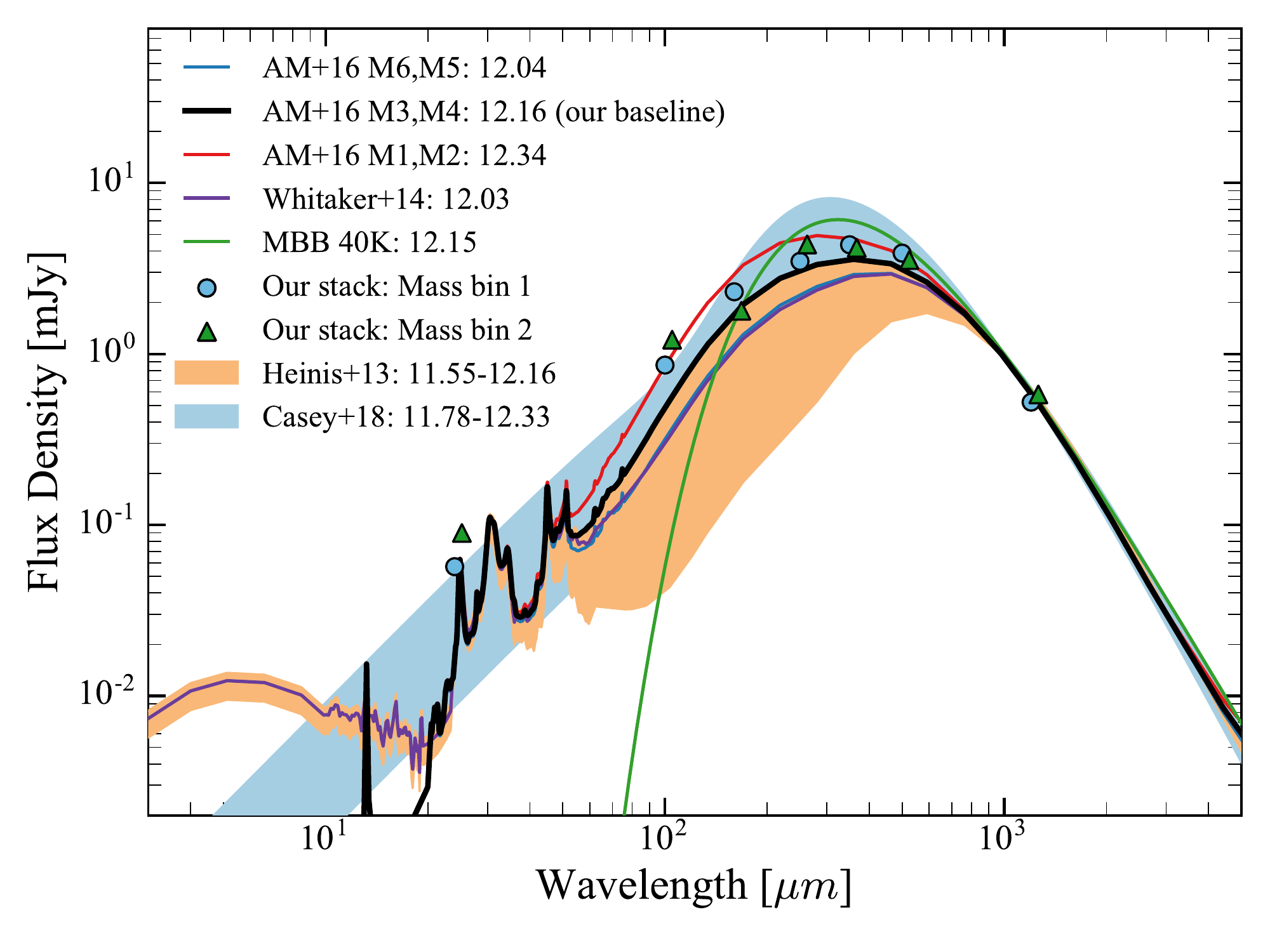}
    \caption{
    Comparison of the IR SED templates used in several previous studies \citep{AlvarezMarquez2016,Heinis2013,Whitaker2014,Casey2018}.
    All templates here are normalised to have $1\,\rm{mJy}$, after redshifting to $z=3.0$.
    For \citet[][denoted as AM+16]{AlvarezMarquez2016}, we plot the three different SED models for the stellar mass ranges  $\rm{log}\,(M_{\ast}/\rm{M_{\odot})=9.75-10.25}$ (M1, M2 model), $\rm{log}\,(M_{\ast}/\rm{M_{\odot})=9.25-10.75}$ (M3, M4 model), and $\rm{log}\,(M_{\ast}/\rm{M_{\odot})=10.75-11.25}$ (M5, M6 model).
    In the upper left legend, we report the IR luminosities obtained from each of the normalised templates in units of $\rm{log}\,(L_{\rm{IR}}/L_{\odot})$.
    The green solid line indicates an optically thin modified black body SED with dust temperature of  $40\,\rm{K}$.
    Most SEDs result in an inferred $\rm{log}\,(L_{\rm{IR}}/L_{\odot})=12.1\pm0.2$, however there are clearly systematic differences between models. For the current work, we adopt a fixed baseline SED: the AM+16 model at $\rm{log}\,(M_{\ast}/\rm{M_{\odot})=9.25-10.75}$, and we add a 0.2 dex systematic uncertainty to all our reported values. 
    For comparison, the stacked {\it Herschel} fluxes for two mass bins of our sample are plotted as triangles ($\rm{log\,(M_{\ast}/M_{\odot})=10.5-11}$) and circles ($\rm{log\,(M_{\ast}/M_{\odot})=11-12}$) after clustering correction, as described in \citet{Magnelli2014}, and after normalising fluxes to match the shown SEDs. 
    }
    \label{fig:SEDcompare}
\end{figure}

In section \ref{sec:IRXmass} we computed the IRX-$M_{\ast}$ relation from our A$^3$COSMOS sample, and compared it to several previous results.
We have found that our IRX-$M_{\ast}$ relation is significantly steeper, such that galaxies with stellar mass of $M_{\ast}<10^{10}\,\rm{M_{\odot}}$ show lower values of IRX than obtained by previous studies. 


However, as pointed out in the text, the shape of the IR SED can significantly affect the inferred $L_{\rm{IR}}$.
Here, we thus investigate the potential effect of using different SEDs. In Figure \ref{fig:SEDcompare}, we plot representative SED templates used in previous studies, that are publicly available, and compare it to our baseline SED model.
This includes SEDs used in \citet{Heinis2013,Whitaker2014,AlvarezMarquez2016,Casey2018}.
For each template, we calculate $L_{\rm{IR}}$ by integrating over $8-1000\,\rm{\mu}m$ after normalising all SED templates to $1\,\rm{mJy}$ at $1000\,\rm{\mu m}$ (i.e. the typical wavelength of our ALMA measurements).

As shown in Fig \ref{fig:SEDcompare}, the $L_{\rm{IR}}$ of our assumed SED template agrees well with that of several other templates within a typical systematic difference of $\sim0.2\,\rm{dex}$.
In principle, this difference in the bolometric correction  depends slightly on the wavelength that is used to normalise. However, the changes are small and do not affect our conclusion.

In \citet{Heinis2013,Whitaker2014,AlvarezMarquez2016}, the authors utilise FIR SED models from \citet{Dale2002}, \citet{Dale2014}, or log averages of \citet{Dale2002}. These SED templates are used to obtain bolometric corrections from stacked $24\,\rm{\mu m}$ fluxes \citep{Whitaker2014}, or to determine the best fit templates to the stacked $250-850\,\rm{\mu m}$ photometry \citep{Heinis2013,AlvarezMarquez2016}.
In \citet{AlvarezMarquez2016}, galaxies are stacked by splitting the sample in six stellar mass bins ($\rm{log}\,(M_{\ast}/\rm{M_{\odot})}=9.75-11.25$ with $\delta \rm{log}\,(M_{\ast}/\rm{M_{\odot})}=0.25$), which resulted in three different SED models. Recall that we use the middle one of these templates appropriate for $\rm{log}\,(M_{\ast}/\rm{M_{\odot})}=10$ as our baseline SED.
As shown in the Figure, most of these SEDs result in similar $L_{\rm{IR}}$ values when normalised to the ALMA flux measurement, with the exception of the templates used in \citet{Heinis2013}, which are somewhat shifted to lower values by up to 0.5dex.

In \citet{Casey2018}, the authors utilise an analytic formula that represents the mid- to far-IR emission of local IR luminous galaxies combining an optically thick modified black body and a mid-IR power-law component.
Additionally, the authors implemented a relationship between $L_{\rm{IR}}$ and peak wavelength ($\lambda_{\rm{peak}}$) to derive their SEDs \citep[Equation (2) of][]{Casey2018b}.
In order to compare to this model, we choose SEDs that have $\lambda_{\rm{peak}}$ corresponding to the range of $L_{\rm{IR}}$ of our sample (i.e. $L_{\rm{IR}}/L_{\odot}=10^{10}-10^{13}$). This large range of SED shapes results in a difference of inferred $\log L_{\rm{IR}}$ of more than 0.5 dex, when normalised to our observed ALMA fluxes. However, the range again spans the $\log L_{\rm{IR}}$ of our baseline SED within $\sim\pm0.2$dex.
Interestingly, using the \citet{Casey2018b} SED model for our analysis would further steepen the IRX-$M_{\ast}$ relation, because $\lambda_{\rm{peak}}$ is set to be inversely proportional to $L_{\rm{IR}}$, resulting in lower dust temperatures at lower IR luminosity or in less massive galaxies. However, for simplicity, we use a fixed SED for all galaxies in the main text. Future observations will be required to test for possible changes in SED shapes with galaxy properties at these redshifts.



The validity of our baseline SED is, however, confirmed by stacks of far-IR fluxes from our galaxy sample. In particular, we performed a stacking analysis of the {\it Herschel} images in two different mass bins using the same method and clustering correction as in \citet{Magnelli2014}. In Figure \ref{fig:SEDcompare}, we show the results for the two highest mass bins, $\rm{log\,(M_{\ast}/M_{\odot})=10.5-11}$ and $\rm{log\,(M_{\ast}/M_{\odot})=11-12}$, for which the Herschel fluxes were well-detected and for which the correction factor for clustering were of order $\sim1.2-1.6$ \citep[see][]{Magnelli2014,Bethermin2015}. For the purpose of this plot, the stacked fluxes were renormalized to fit the plotted SEDs. As can be seen, the shape of our baseline FIR SED is well-matched to the observed fluxes.





The above analysis suggest that the use of our baseline SED template from \citet{AlvarezMarquez2016} is well justified and that our derived $L_{\rm{IR}}$ values are consistent with a range of different SED models within a systematic uncertainty of $\sim\pm0.2$dex, which we add to our derived values in the main text of the paper.



\bsp	
\label{lastpage}
\end{document}